\documentclass{mn2e}
\usepackage{epsf,times}
\newcommand{\etal}{{et al}\/.}
\newcommand{\hh}{^{\rm h}}
\newcommand{\mm}{^{\rm m}}

\begin{document}
\title[{\rm Chandra} observation of 3C\,31]{A {\it Chandra} observation of the X-ray environment and jet of 3C\,31}
\author[M.J.~Hardcastle \etal]{M.J.\ Hardcastle$^1$, D.M.\ Worrall$^1$,
M.\ Birkinshaw$^1$, R.A.\ Laing$^{2,3}$ and A.H.\ Bridle$^4$\\
$^1$ Department of Physics, University of Bristol, Tyndall Avenue,
Bristol BS8 1TL\\
$^2$ Space Science and Technology Department, CLRC, Rutherford
Appleton Laboratory, Chilton, Didcot, Oxfordshire OX11 0QX\\
$^3$ University of Oxford, Department of Astrophysics, Nuclear and
Astrophysics Laboratory, Keble Road, Oxford OX1 3RH\\
$^4$ National Radio Astronomy Observatory, 520 Edgemont Road,
Charlottesville, VA 22903-2475, U.S.A\\
}
\maketitle
\begin{abstract}
We have used a deep {\it Chandra} observation of the central regions of the
twin-jet FRI radio galaxy 3C\,31 to resolve the thermal X-ray
emission in the central few kiloparsecs of the host galaxy, NGC 383, where the
jets are thought to be decelerating rapidly. This allows us to make
high-precision measurements of the density, temperature and pressure
distributions in this region, and to show that the X-ray emitting gas
in the centre of the galaxy has a cooling time of only $5 \times 10^7$
years. In a companion paper these measurements are used to place
constraints on models of the jet dynamics.

A previously unknown one-sided X-ray jet in 3C\,31, extending up to 8
arcsec from the nucleus, is detected and resolved. Its structure and
steep X-ray spectrum are similar to those of X-ray jets known in other
FRI sources, and we attribute the radiation to synchrotron emission
from a high-energy population of electrons. {\it In situ} particle
acceleration is required in the region of the jet where bulk
deceleration is taking place.

We also present X-ray spectra and luminosities of the galaxies in the
Arp 331 chain of which NGC 383 is a member. The spectrum and
spatial properties of the nearby bright X-ray source 1E~0104+3153 are
used to argue that the soft X-ray emission is mostly due to a foreground
group of galaxies rather than to the background broad absorption-line
quasar.
\end{abstract}

\section{Introduction}

The radio emission from low-luminosity radio galaxies [class
I of Fanaroff \& Riley (1974), hereafter FRI] is often dominated 
by bright jets which may extend for many hundreds of kpc on 
both sides of the central active nucleus.  A number of arguments,
including observations of superluminal motion in some sources,
lead to the conclusion that the jet material moves at
relativistic bulk speeds on parsec scales.  At large distances
from the galaxy, however, the diffuse, relaxed appearance of 
the jets, their large opening angles, and their brightness
symmetry across the nucleus (together with other indicators) 
suggest that they are trans-sonic or subsonic flows with
sub-relativistic speeds. The jets must therefore decelerate
on intermediate scales.

There is now a good deal of direct evidence, both from studies of
Doppler beaming of individual objects (e.g.\ Hardcastle \etal\ 1997;
Laing \& Bridle 2002a) and samples (Laing \etal\ 1999) and from
independent indicators of source orientation such as depolarization
(Morganti \etal\ 1997) that this deceleration largely happens on
scales of $\sim 1$--$10$ kpc from the nucleus, corresponding to scales
of a few arcseconds at the typical distances of well-studied objects.
A plausible mechanism for deceleration is mass loading by
the entrainment of thermal material, either by injection 
from stellar winds within the volume that is traversed by
the jet (e.g., Bowman, Leahy \& Kommissarov 1996) or by
ingestion from the galactic atmosphere across a turbulent 
boundary layer at the edges of the jet (e.g., Bicknell 1984).
All models in which the jet decelerates rapidly by mass loading 
require the presence of an external pressure gradient to
prevent disruption of the jet.

3C\,31 is a nearby twin-jet FRI radio galaxy ($z=0.0169$), hosted by
the D galaxy NGC 383, a member of the optical chain Arp 331. 3C\,31's
bright pair of jets was among the first to be studied in detail at
radio frequencies (e.g.\ Burch 1977; Fomalont \etal\ 1980). An optical
counterpart to the brighter northern jet, first discovered by Butcher,
van Breugel \& Miley (1980), was not seen in later observations by
Keel (1988) and Fraix-Burnet (1991) but appears to be real (Croston
\etal\ in preparation). Recent high-quality radio imaging with the
Very Large Array (VLA) allowed Laing \& Bridle (2002a) to make
detailed models of the velocity field in the jets within 30 arcsec of
the nucleus, on the assumption that the jets are intrinsically
symmetrical and anti-parallel, so that apparent differences between
the brightness and polarization structures arise entirely from
differences in Doppler beaming and aberration between approaching and
receding flows. They inferred the angle to the line of sight to be
$52\degr$ with an uncertainty of a few degrees, and found that the
jets decelerate from on-axis speeds of $v/c \approx 0.9$ at 1 kpc from
the nucleus to $v/c \approx 0.22$ at 12 kpc along the jet, with slower
speeds at the jet edges.
\label{optjet}

In order for the flaring and recollimation observed in 3C\,31 to occur
over distances of 1 -- 4 kpc, there must be a substantial pressure
gradient on this scale. This can only be provided by gas in the hot,
X-ray emitting phase. X-ray emission associated with 3C\,31 has
previously been observed with {\it Einstein} (Fabbiano \etal\ 1984)
and with the {\it ROSAT} PSPC and HRI (Trussoni \etal\ 1997; Komossa
\& B\"ohringer 1999; Canosa \etal\ 1999; Hardcastle \& Worrall 1999).
Extended X-ray emission from the rich host group of NGC 383 has been
seen on arcminute scales, and is well modelled as a thermal
plasma with a temperature $1.5 \pm 0.1$ keV. Komossa \& B\"ohringer
fitted it with a $\beta$ model with $\beta = 0.38$ and core radius 154
arcsec. In addition, all the elliptical galaxies in the chain,
including NGC 383 itself, are X-ray sources. Komossa \& B\"ohringer
carried out a spectral fit to the PSPC observations of NGC 383 and
obtained good fits with a combination of a power law (from the active
nucleus) and weaker thermal emission with $kT \sim 0.6$ keV (from
unresolved hot gas centred on the galaxy). However, the spatial
resolution of the {\it ROSAT} PSPC, or even the HRI, was insufficient
to resolve thermal emission on scales matched to the region of strong
deceleration in the jet or to separate it from non-thermal X-ray
emission associated with the active galaxy.

In this paper we present a deep {\it Chandra} observation of the
nuclear regions of 3C\,31, taken with the aim of accurately
constraining the pressure gradient in the inner few arcseconds. We
also discuss a newly discovered X-ray jet in 3C\,31, and the X-ray
emission from the galaxies in the Arp 331 chain and the background
source 1E 0104+3153. A companion paper (Laing \& Bridle 2002b)
presents an analysis of the dynamics of the jets in 3C\,31 using the
kinematic model of Laing \& Bridle (2002a), the external density and
pressure distributions from our {\em Chandra} observations, and a
conservation-law approach based on that of Bicknell (1994). There it
is shown that the jets can indeed be decelerated by mass injection and
recollimated by the external pressure gradient, and the variations of
internal pressure and density along them are inferred.

Throughout the paper we use a cosmology with $H_0 = 70$ km s$^{-1}$
Mpc$^{-1}$ and $q_0 = 0$, so the linear scale is 340 pc arcsec$^{-1}$ at
the redshift of the host galaxy.

\section{The observations}

\begin{figure*}
\epsfxsize 16cm
\epsfbox{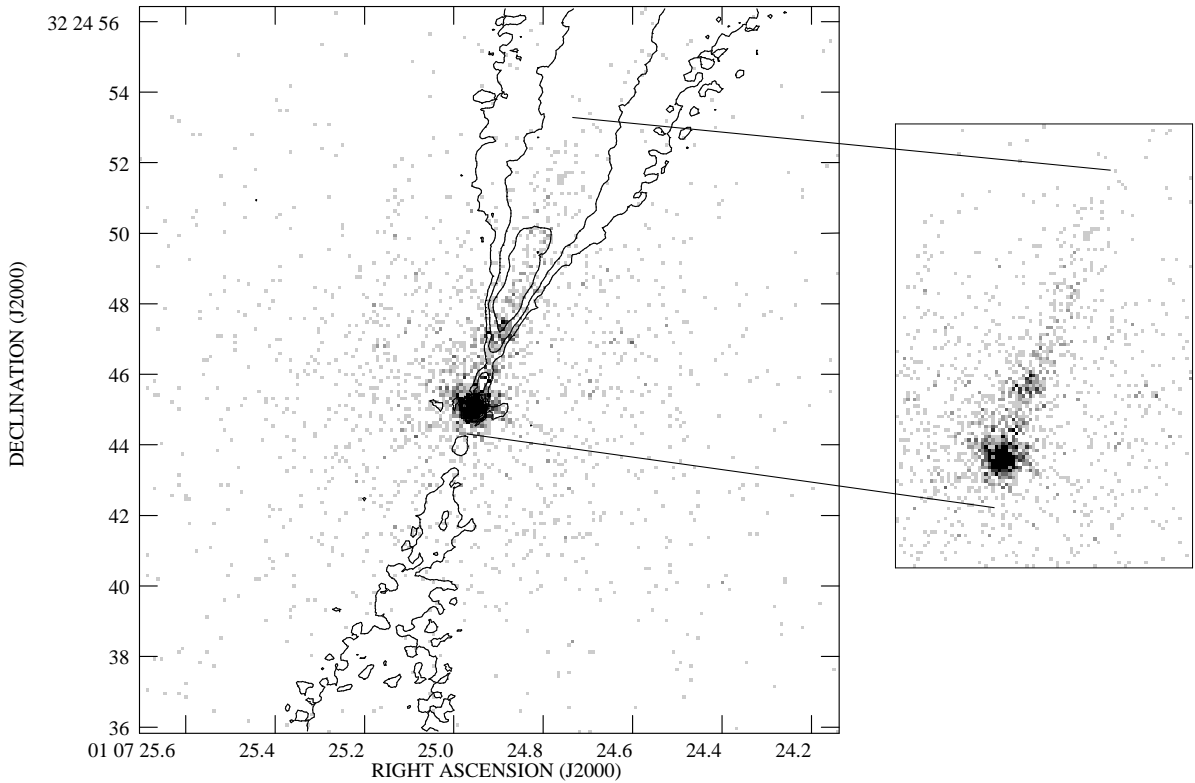}
\caption{{\it Chandra} image of the inner regions of 3C\,31. The
greyscale has a pixel size of 0.0946 arcsec, with black being 5 counts
per pixel. Left: superposed are contours from an 8.5-GHz radio map at a
resolution of 0.25 arcsec, at $20 \times (1, 4, 16, \dots)$ $\mu$Jy
beam$^{-1}$. The core X-ray centroid has been aligned with the radio
core position as discussed in the text. Right: the X-ray jet alone, on
the same scale.}
\label{hijet}
\end{figure*}

We observed 3C\,31 for 44410 s with the {\it Chandra} ACIS-S on 2000
November 6th. 3C\,31 was positioned near the aim point on the
back-illuminated S3 CCD. There were no high-background intervals or
aspect excursions during the observation, so no temporal filtering of
the data was required. We reprocessed the observation using {\sc ciao}
version 2.1 to generate a new level 2 events file containing events
with the standard set of event grades but with the 0.5-pixel
randomization removed.

Fig.\ \ref{hijet} shows an image of the inner regions of 3C\,31, made
from all events with energies between 0.5 and 7.0 keV. (Throughout the
paper, images and spectral fits use the events in this energy range.)
Extended emission on scales of the host galaxy, and an X-ray core and
jet, are detected. Their spectral and spatial properties are discussed
in the following sections. Spectra were extracted and corresponding
response matrices constructed using {\sc ciao}, and model fitting was
carried out in {\sc xspec}. In all cases the extracted spectra were
grouped so that each fitting bin had $>20$ counts. Errors quoted
throughout are the $1\sigma$ value for one interesting parameter,
unless otherwise stated. All power-law indices are the energy index or
spectral index, $\alpha$, defined in the sense that the flux density
$S \propto \nu^{-\alpha}$.

\section{The X-ray core}
\label{core}

In an extraction circle with radius 1.5 arcsec centred on the X-ray
nucleus of 3C\,31, using a source-centred background annulus with
radii between 2 and 5 arcsec (excluding jet emission), there are $835
\pm 33$ net counts. The spectrum of this emission is very poorly
fitted either with a single power-law model ($\chi^2/n = 84/32$,
$\alpha = 1.6 \pm 0.2$, $N_H = (1.8 \pm 0.4) \times 10^{21}$
cm$^{-2}$) or with a thermal ({\sc mekal}) model ($\chi^2/n = 90/31$,
$kT = 0.98 \pm 0.11$, $N_H = (1.4 \pm 0.3) \times 10^{21}$ cm$^{-2}$,
abundance 0.04). It is better fitted ($\chi^2/n = 22/30$) with a
combination of the two: fixing the absorbing column for the thermal
component at the Galactic value ($5.53 \times 10^{20}$ cm$^{-2}$) and
the abundance to 0.2 solar (see section \ref{gas}), the best-fitting
power-law model has an intrinsic absorbing column at the redshift of
the galaxy of $(1.5_{-1.4}^{+2.2}) \times 10^{21}$ cm$^{-2}$ and a
power-law index of $0.52_{-0.16}^{+0.29}$, while the thermal component
has $kT = 0.66 \pm 0.04$ keV and contributes $55 \pm 10$ per cent of the
observed emission at 1 keV, or $450 \pm 80$ net counts.  The unabsorbed 1-keV
flux density of the power-law component of the core in this model is
$9 \pm 2$ nJy. The radio-to-X-ray spectral index, $\alpha_{\rm
RX}$, between the arcsecond-scale nuclear radio emission (using the
8.4-GHz flux density of 91 mJy) and the power-law component of the
core X-ray emission is $0.94 \pm 0.01$, which is similar to the values
for the B2 radio sources observed by Worrall \etal\ (2001).

When we use an extraction region matched to that of Canosa \etal\
(1999), who determined a flux density of 64 nJy for the nuclear region
of the source from {\it ROSAT} HRI data, we find that the {\it
Chandra} X-ray flux in the {\it ROSAT} band is consistent. There is
thus no evidence for X-ray variability of 3C\,31's nucleus, though
this is a weak constraint; variability at the level of tens of per
cent in the net count rate would not have been detectable given the
{\it ROSAT} uncertainties, so that the non-thermal component of the
core could have varied by $\sim 50$ per cent between the {\it ROSAT}
and {\it Chandra} observations. The larger {\it ROSAT} flux density
quoted by Canosa \etal\ includes emission from the thermal component
of the core, from the X-ray jet (Section \ref{jet}) and from the
extended thermal emission around 3C\,31 (Section \ref{gas}).

The lack of strong absorption of the power-law component of 3C\,31
(there is no more absorption than would be expected given the observed
dust features in {\it Hubble Space Telescope} images) is consistent
with the results we have obtained for other FRI radio galaxies
(Worrall \etal\ 2001, Hardcastle \etal\ 2001). These results imply
that the line of sight to the nuclear power-law component in these
objects does not intersect the kind of `torus' inferred from
observations of some FRII radio sources (e.g., Ueno et al. 1994).
Similar considerations apply to the nuclear optical emission observed
from the host galaxies of FR\,I sources (Chiaberge et al.\ 1999,
Hardcastle \& Worrall 2000). Two possibilities can be envisaged: a)
the nuclear optical and X-ray emission originates mainly in the jet on
scales larger than the torus, and is therefore not absorbed; or b)
there is no torus, in which case some or all of the observed X-ray
emission could originate in the active nucleus itself (from the
accretion disc and environs). We have argued elsewhere (e.g., Canosa
\etal\ 1999; Hardcastle \& Worrall 2000) for the former possibility,
based on the observed correlation between radio and X-ray core
emission, which suggests that the nuclear soft X-ray emission
originates in the jets, presumably on pc- to 100-pc scales.

However, we note that infra-red observations have led Whysong \&
Antonucci (2001) and Perlman \etal\ (2001) to infer that there is no
torus (or, more generally, no heavily absorbed nuclear component above
a certain bolometric luminosity) in one FRI radio galaxy, M87. Most
Chandra observations of FRI sources do not show any evidence for a
heavily absorbed nuclear X-ray component, although Birkinshaw \etal\
(in preparation) will report on a detection of an absorbed nuclear
component in NGC~4261. X-ray observations are therefore consistent
with the absence of a torus in some sources, in turn implying that we
cannot rule out the possibility of a contribution from the active
nucleus to the observed nuclear X-ray emission in 3C\,31. (Jet-related
X-rays must still dominate in a substantial fraction of the FR\,I
population to explain the observed radio/X-ray correlation.) If the
X-rays from 3C\,31's nucleus {\it do} originate in the small-scale
jet, then their flat spectrum is most consistent with an
inverse-Compton model, as predicted by Hardcastle \& Worrall
(2000). However, the other {\it Chandra} evidence on this point is
inconsistent: M87 (Wilson \& Yang 2002), 3C\,66B (Hardcastle \etal\
2001) and the three objects studied by Worrall \etal\ (2001) have all
exhibited steeper nuclear spectra, $\alpha \approx 1$, though in some
cases the errors were large. We defer detailed discussion of the radio
to X-ray nuclear spectra of FRI sources to a future paper.

Using the default {\it Chandra} astrometry, the core centroid is
displaced by about 0.13 arcsec north of the accurately known
radio core position. This offset is within the known uncertainty in
{\it Chandra} source positions, so we have corrected the X-ray
positions to remove the displacement. However, since there is a
substantial contribution to the X-ray counts in the nuclear region
from thermal emission which may not be centred on the active nucleus,
the accuracy of the alignment between the radio and X-ray frames is
probably no better than about 0.05 arcsec.

\section{The X-ray jet}
\label{jet}
X-ray emission is clearly detected from the brighter, northern radio
jet of 3C\,31, extending about 8 arcsec from the nucleus. Fig.\
\ref{hijet} shows that most of this emission corresponds to the
well-collimated inner region of the jet at radio wavelengths, with the
peak intensity in the X-ray emission occurring where the radio
emission begins to brighten and decollimate. A rectangular region (6.5
arcsec by 1.7 arcsec) around the jet but avoiding the core, with
background taken from adjacent identical rectangles, contains $368 \pm
35$ net counts. A power-law model of the spectrum, assuming Galactic
absorption, gives a good fit to the data ($\chi^2/n = 7.3/13$) with
$\alpha = 1.09 \pm 0.16$. The improvement in the fit produced by
allowing the absorbing column to vary is not significant on an F-test.
The spectrum can also be fitted adequately ($\chi^2 = 11.5/13$) with a
thermal model with $kT = 2.3_{-0.3}^{+0.7}$ keV, Galactic absorption
and a (fixed) abundance of 0.2 solar. On the preferred power-law model, the
unabsorbed flux density of the jet at 1 keV is $7.3 \pm 0.5$ nJy.

\begin{figure*}
\hbox{\epsfxsize 8.0cm
\epsfbox{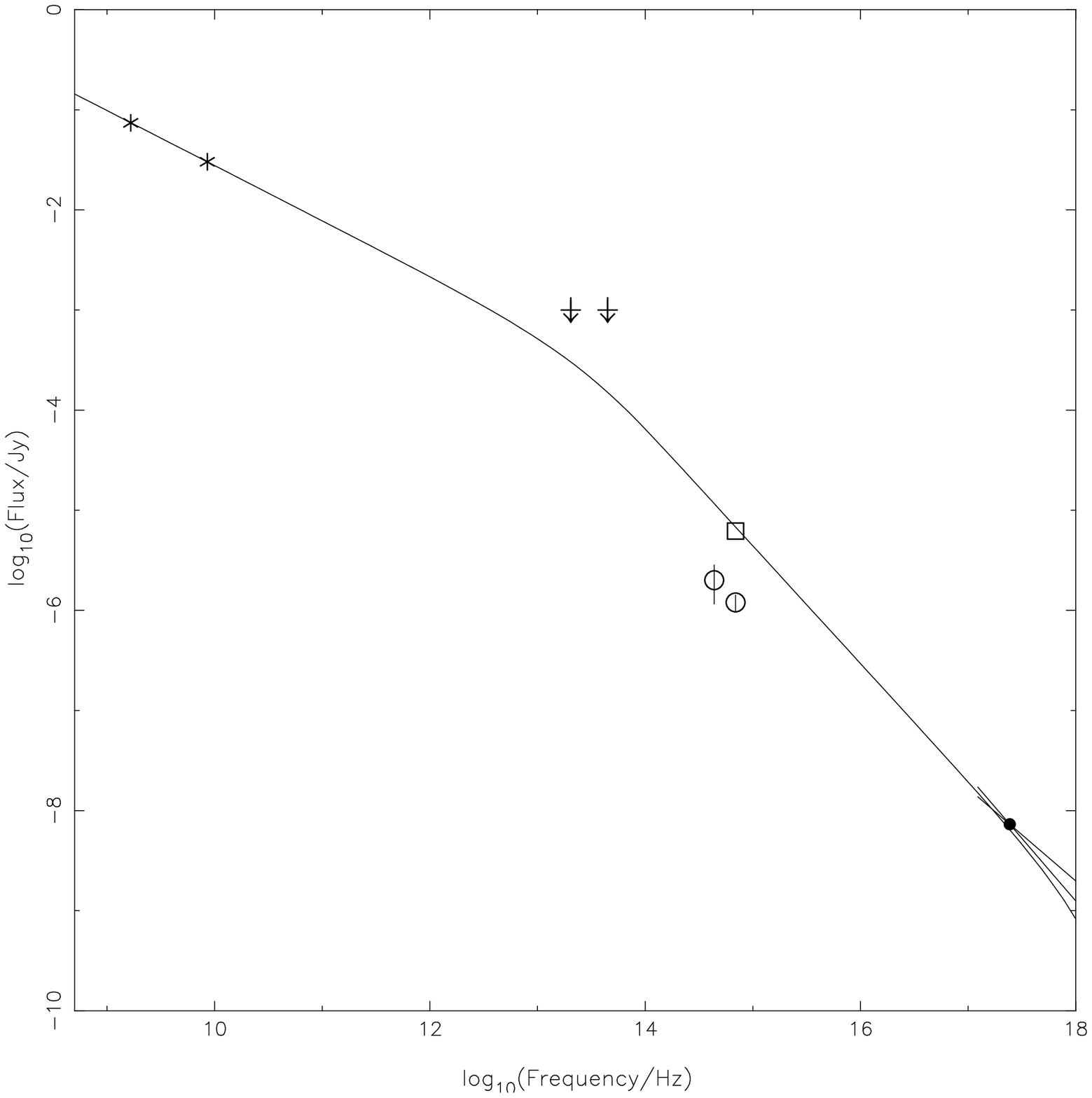}
\hskip 5pt
\epsfxsize 8.0cm
\epsfbox{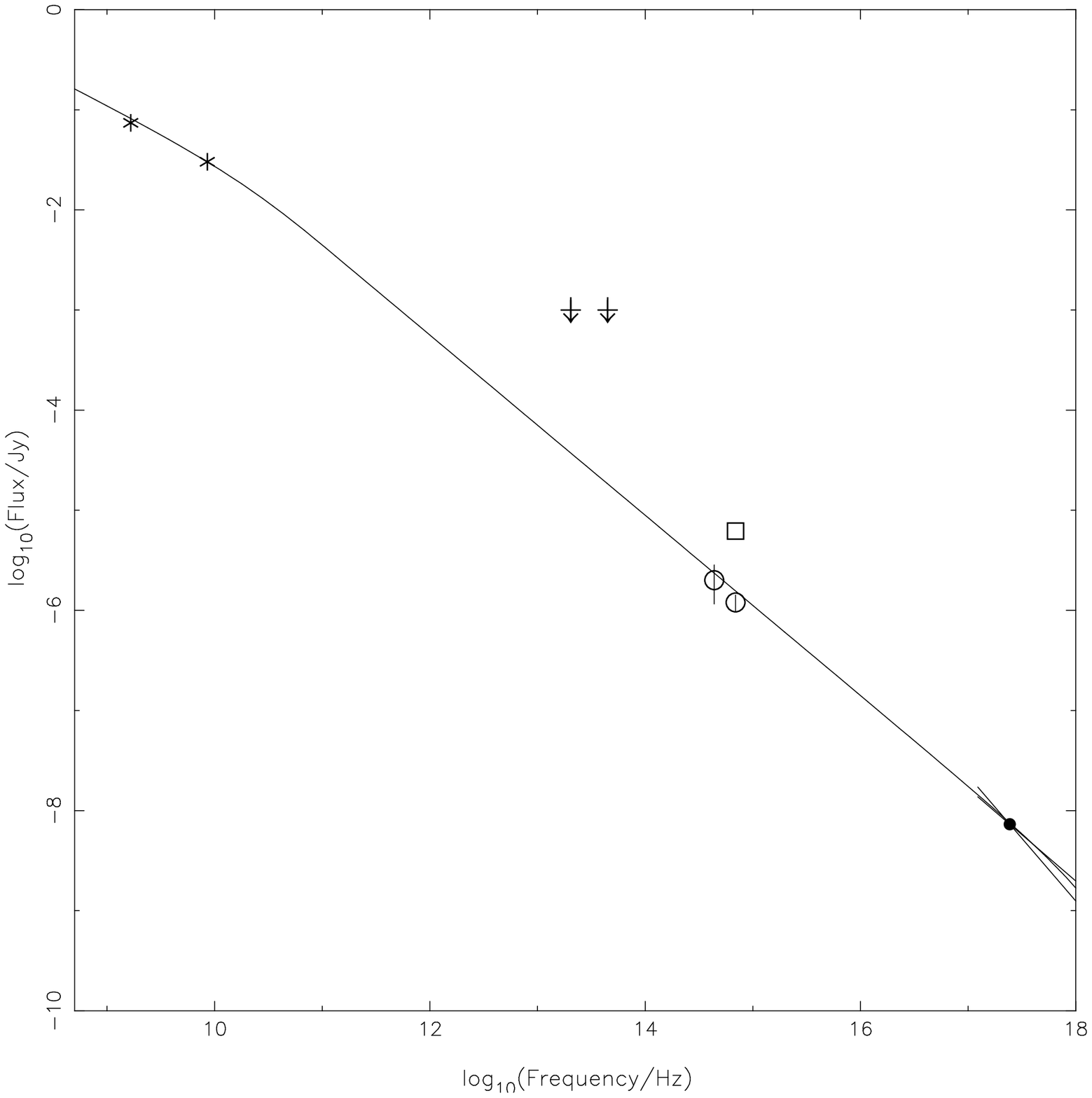}}
\caption{The radio-to-X-ray spectrum of the jet of 3C\,31, based on
the whole 6-arcsec region of the jet detected with {\it Chandra}.
Radio points (stars) are taken from the data of Laing \& Bridle
(2002a); the infra-red data points (crosses) are upper limits derived
from our ISO observations (Tansley \etal\ 2002); the high optical data
point (box) is the flux density of Butcher \etal\ (1980), while the
lower optical data points (open circles) are the data of Croston
\etal\ (in preparation); the filled circle is the X-ray flux density
reported in the text. The `bow tie' around the X-ray flux density
illustrates the $1\sigma$ range of spectral index in the power-law fit
to the jet spectrum. The solid lines show broken power-law synchrotron
models of the kind discussed in the text: left, the model fitting the
Butcher \etal\ optical flux, with a spectral break in the infra-red;
right, the model fitting the Croston \etal\ optical fluxes, with a
break in the mm-wave region.}
\label{netspec}
\end{figure*}

Two X-ray jets in FRIs, in the nearby objects M87 (Harris \etal\ 1997)
and Centaurus A (Turner \etal\ 1997) were known before the launch of
{\it Chandra}, but {\it Chandra} observations have resulted in the
detection of a number of new FRI X-ray jets, including those in B2
0206+35 and B2 0755+37 (Worrall, Birkinshaw \& Hardcastle 2001) and
3C\,66B (Hardcastle, Birkinshaw \& Worrall 2001). In the latter paper
we argued that the X-ray emission from all known FRI X-ray jets was
consistent with being synchrotron emission from the high-energy tail
of the electron population responsible for the radio emission. The
radio-to-X-ray ratio and X-ray spectral index of 3C\,31's jet are very
similar to those observed in other sources, suggesting that a
synchrotron model can be applied to 3C\,31 as well. The steep spectrum
of the X-ray emission would not be expected in an inverse-Compton
model for the jet emission, and for magnetic fields close to
equipartition standard inverse-Compton models (where the scattered
photon population is the synchrotron photons or the CMB) underestimate
the observed X-ray flux density by three to four orders of magnitude.
The well-constrained jet speeds and angle to the line of sight (Laing
\& Bridle 2002a) mean that beamed inverse-Compton models (e.g.\
Tavecchio \etal\ 2000) do not apply here. As we argued in the case of
3C\,66B, light from the hidden BL Lac object in the core of the AGN
probably dominates the photon density in the jet. In this case, we
could produce the observed X-ray flux density (but not the spectrum)
if the hidden BL Lac (whose spectrum is modelled here as a simple
power law from radio through to optical) had an effective flux
density, as seen from the jet direction, of about 1000 Jy in the
radio. This is very large, and requires high beaming Lorentz factors
($\sim 7$) in the presumed BL Lac nuclear jet, at $52\degr$ to the
line of sight, in order not to over-produce the observed radio core
emission. For more plausible values of the hidden BL Lac flux density,
substantial departures from equipartition are still required to
produce the observed X-ray emission by the inverse-Compton mechanism,
and the departures become even larger if the expected relativistic
motion of the jet away from the nucleus is incorporated in the models.
Because of this, and of the steep X-ray spectrum, we favour a
synchrotron model for the X-ray emission from the jet.

The 3C\,31 jet's optical properties are less well known than
in M87 and 3C\,66B. As discussed in Section \ref{optjet}, several
studies have failed to detect any optical emission at all, while
Butcher \etal\ (1980) claimed a significant detection. Large
uncertainties arise because of the central dust lane and the
bright rim of stellar emission around it (Martel \etal\ 1999), which
unfortunately cross the brightest part of the radio jet. Most
recently, Croston \etal\ (in prep.) have measured a flux density for
the optical emission {\it outside} this rim structure. If we scale
up their flux to match our X-ray extraction region, using the ratio of
radio flux densities inside and outside the X-ray region to account
for the optical emission obscured by the dust lane and its rim, we
obtain net flux densities in the $R$ and $B$ bands of $2.0 \pm 0.8$
and $1.2 \pm 0.3$ $\mu$Jy respectively, although this scaling procedure
assumes that there is no variation in radio-optical spectral index
along the jet. As Fig.\ \ref{netspec} shows, the uncertainties in the
optical flux lead to uncertainties in the best-fitting synchrotron
model: if we adopt the value of Butcher \etal\ (1980) we can fit a
model which is quite similar to that seen in M87 and 3C\,66B, in
which the spectrum breaks from its radio value of $\alpha = 0.55$
somewhere in the infra-red by $\Delta\alpha = 0.63$, whereas if we use
the values of Croston \etal\ we require a smaller spectral break
($\Delta\alpha = 0.35$) which occurs at much lower frequency in order
to accommodate the X-ray, optical and radio data.

Although the SED is currently uncertain, a synchrotron model of some
sort is certainly viable for 3C\,31. Given the very short synchrotron
lifetimes of the X-ray emitting electrons, such a model requires {\it
in situ} acceleration of at least the high-energy electron
population. One possible source of energy for particle acceleration,
at least in the outer regions of the X-ray jet, is the strong bulk jet
deceleration inferred by Laing \& Bridle (2002a).

\begin{figure}
\epsfxsize \linewidth
\epsfbox{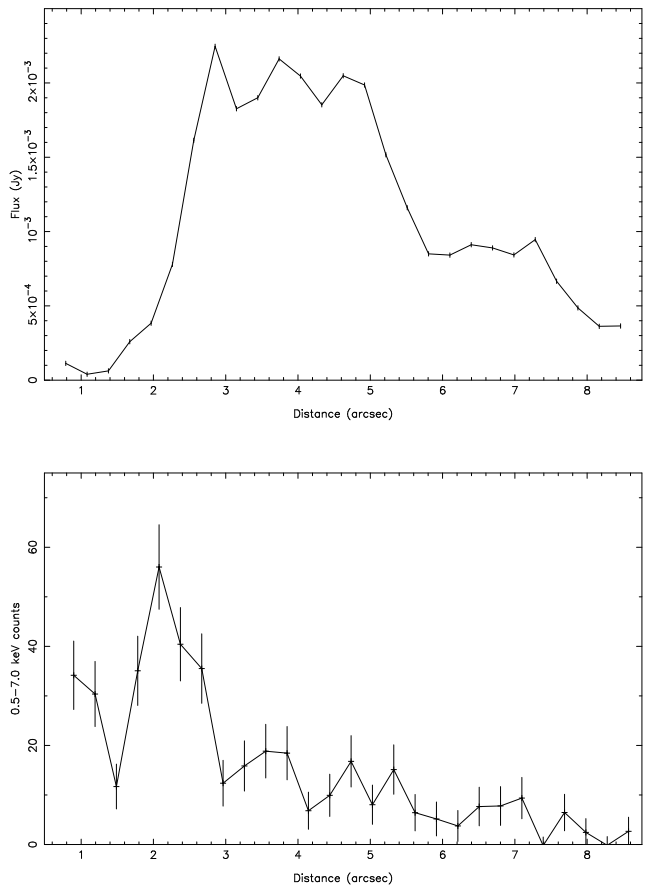}
\caption{Profiles of the jet in the radio (upper panel) and X-ray
(lower panel). Profiles were made using the maps of Fig.\ \ref{hijet},
extracting all the emission from a series of rectangles of 0.3 arcsec
by 2.0 arcsec (long axis perpendicular to the jet axis). Background
subtraction was carried out using adjacent rectangles for each bin.
The innermost two bins of the X-ray profile are dominated by emission
from the X-ray nucleus. Errors on the points on the X-ray
profile are $1\sigma$; note that the X-ray data points are not
independent, since the sampling along the jet axis is smaller than the
resolution of {\it Chandra} ($\sim 0.6$ arcsec FWHM). The restoring
beam used in the radio maps was $0.25$ arcsec FWHM, so that the radio
points are effectively independent. However, the errors on these
points are $1\sigma$ based on the off-source noise, and so are lower
limits on the actual errors.}
\label{slice}
\end{figure}

As Fig.\ \ref{hijet} shows, both the X-ray and radio emission from the
jet are comparatively smooth, with little sign of the knotty structure
seen in M87 or 3C\,66B. However, there is a clear X-ray peak at just
over 2 arcsec from the core, containing about 100 counts in total,
which lies in the region where the jet's radio flux increases steeply
as a function of distance from the core. The knot can be seen in the
X-ray profile along the jet shown in Fig.\ \ref{slice} and is
associated with the well-collimated inner region defined by Laing \&
Bridle (2002a). In 3C\,66B, Centaurus A and M87 (Hardcastle \etal\
2001; Kraft \etal\ 2002; Wilson \& Yang 2002) unexpectedly strong
X-ray emission is seen from the radio-weak inner parts of the jet, and
it seems plausible that we are seeing a similar phenomenon in 3C\,31.
In the case of 3C\,66B, we argued that a difference in the particle
acceleration process (or conceivably a different emission mechanism)
was required in the inner jet. (This sort of offset between radio and
X-ray peaks is expected in a nuclear inverse-Compton model, but such
models are not favoured here for the reasons discussed earlier.) The
flow in the inner 3 arcsec of the jet is inferred by Laing \& Bridle
(2002a) to have a fast component ($v/c \approx $ 0.8 -- 0.9) together
with much slower material, although the transverse velocity profile is
unresolved; there are therefore at least two possible components of
the jet from which this excess X-ray emission could originate. It is
tempting to relate the excess in some way to the reconfinement shock
at which the internal pressure of the jet falls below the pressure of
the external medium, which, as discussed by Laing \& Bridle (2002b), is
expected to occur somewhere in this region. But in the absence of a
detailed model for the velocity structure in this faint part of the
jet we cannot give a definitive explanation for this component.

\begin{figure*}
\hbox{\epsfxsize 8.0cm
\epsfbox{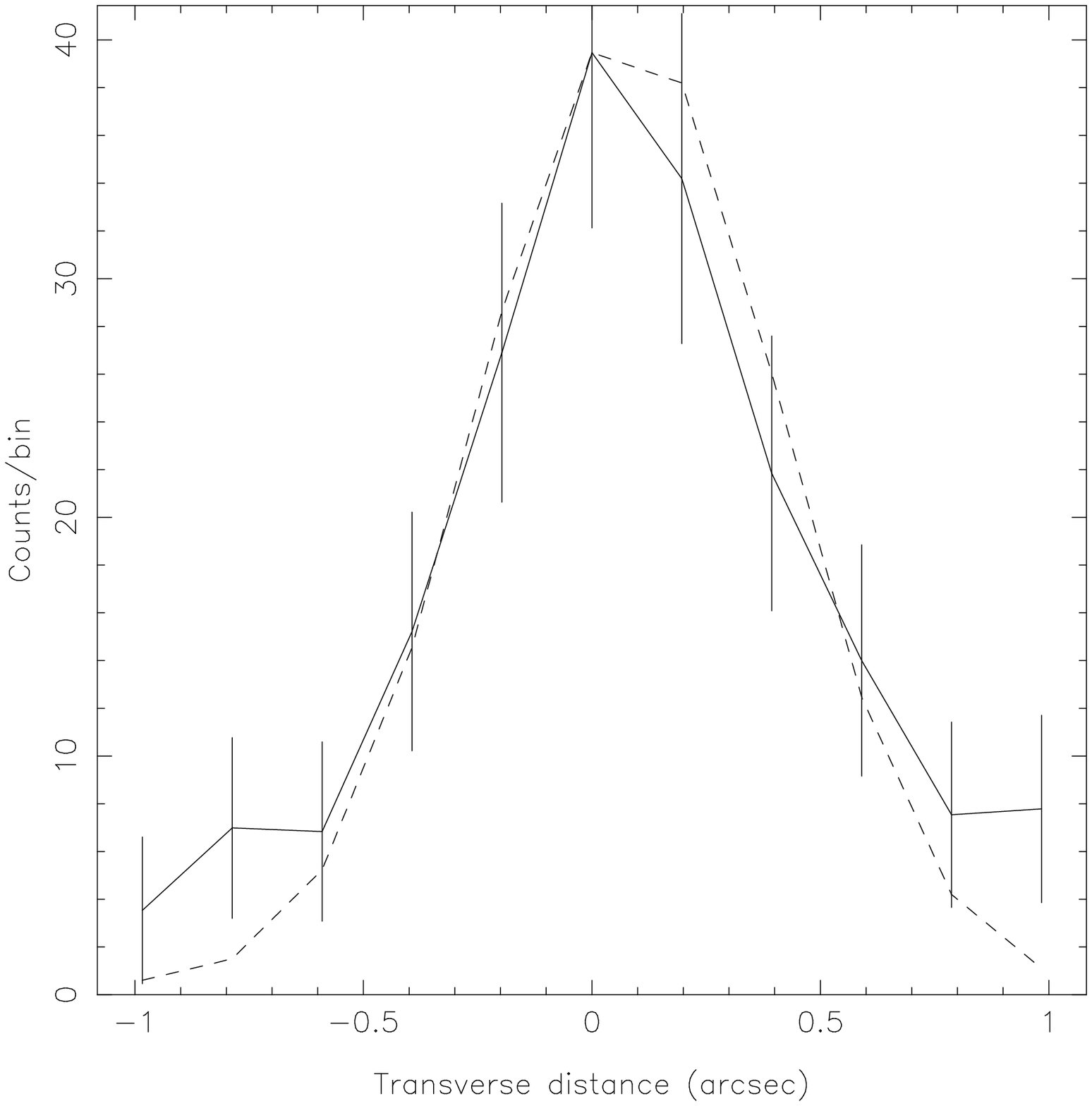}
\hskip 5pt
\epsfxsize 8.0cm
\epsfbox{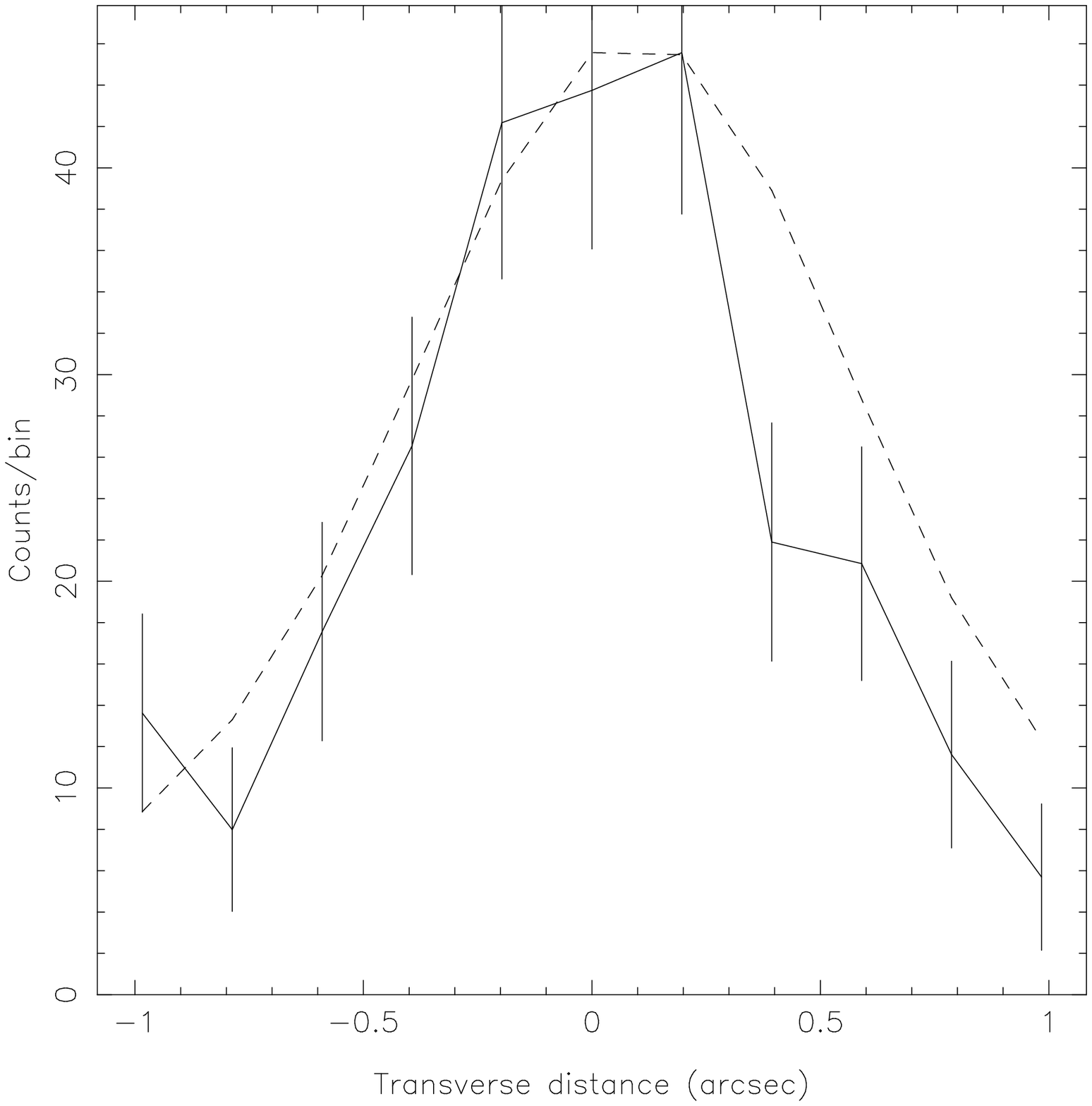}}
\caption{Transverse slices across the jet in X-ray (solid line) and
radio (dashed line). Left: the inner jet (1.5 to 2.5 arcsec from the
nucleus). Right: the flaring region of the jet (2.5 to 8 arcsec from
the nucleus).  Each bin represents a rectangle 0.2 arcsec wide. The
zero of the abscissa is on the ridge line of the jet and the positive
direction is to the SW. The radio data, from an 8.4-GHz image
convolved to a resolution of 0.65 arcsec (roughly matching the {\it
Chandra} resolution), is plotted for comparison, scaled so that the
maxima of the two curves are the same.}
\label{tslice}
\end{figure*}

It is important to note that optical and X-ray jet emission is also
associated with the flaring region defined by Laing \& Bridle (2002a),
between 2.5 and 8.2 arcsec from the nucleus, where deceleration is
inferred to start. No X-ray emission is detected from the outer region
($>8.2$ arcsec from the nucleus). In the flaring region, the
radio-to-X-ray ratio appears constant within the errors. There is no
evidence for a spectral difference between the inner X-ray jet ($<2.5$
arcsec) and the outer jet ($>2.5$ arcsec), though the errors are
large. Nor is there any radio spectral difference between the flaring
region of the jet and the rest. Transverse profiles taken across the
inner and flaring regions (Fig.\ \ref{tslice}) also show little
difference between them; there is no strong evidence that either of
the regions is more centrally peaked in the X-ray than in the radio.

No X-ray emission is detected from the counterjet. The radio
jet-to-counterjet ratio, $R_{\rm radio}$, is $\sim 10$ in the inner 6
arcsec, so we might have expected to see $\sim 40$ counts from the
counterjet if the ratio were the same in the X-ray. In a relativistic
beaming model for the synchrotron radiation, however, the steeper
spectrum of the X-ray emission gives rise to a greater K-correction,
so that the X-ray and radio jet-counterjet ratios should not be the
same. The predicted $R_{\rm X}$ is $R_{\rm radio}^{(2+\alpha_{\rm
X})/(2+\alpha_{\rm radio})}$, or about 16, leading to a prediction of
about 25 X-ray counts in the X-ray counterjet. Our $3\sigma$ upper
limit on X-ray counts from the the counterjet region is about 30
counts. The jet-to-counterjet ratio in the X-ray is therefore
consistent with relativistic beaming of an intrinsically identical
synchrotron X-ray jet and counterjet; there is no evidence for an
intrinsic difference between the X-ray emission on the jet and
counterjet sides.

\section{The chain galaxies}

\begin{figure*}
\epsfxsize 10cm
\epsfbox{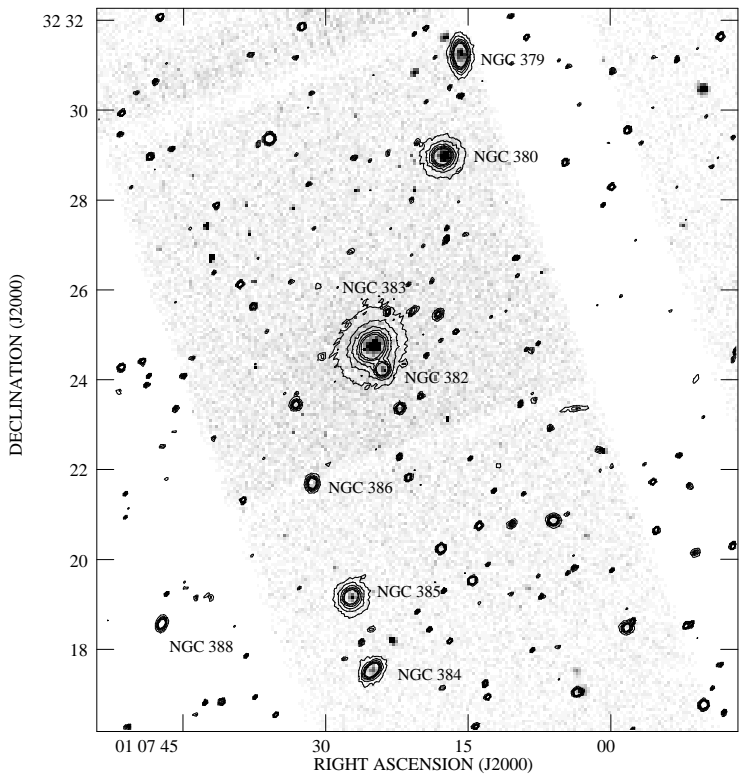}
\caption{{\it Chandra} image of the galaxies in the Arp 331 chain;
pixels are 3.94 arcsec on a side.
Overlaid are contours from the red Digital Sky Survey-2 plates. No
exposure correction has been applied, and {\it
Chandra} chip boundaries are visible; NGC 379, 380, 383 (3C\,31) and
382 lie on the S3 chip, while NGC 385 and 384 lie on the
front-illuminated S2 chip.}
\label{galaxies}
\end{figure*}

All the bright galaxies of the Arp 331 chain, including NGC 383's companion
galaxy NGC 382, are detected in the {\it Chandra} observation (Fig.\
\ref{galaxies}). We have extracted spectra for each object using
12-arcsec source circles centred on the galaxy, with background taken
from annuli between 12 and 15 arcsec; we use matched source regions to
facilitate comparisons between sources. The resulting count rates and
spectral fits are tabulated in Table \ref{companions}. We 
tried to fit a power-law with Galactic absorption or a {\sc mekal}
model with Galactic absorption and abundance 0.2 solar to each spectrum.
If neither model gave a satisfactory fit we allowed the absorbing
column or the abundance to vary, or fitted a model combining thermal
and power-law components. We tabulate only satisfactory fits,
with $\chi^2/n \la 2$.

\begin{table*}
\caption{X-ray data for the Arp 331 galaxies}
\label{companions}
\begin{tabular}{lrp{3.3cm}lrr}
\hline
Source&Count rate&Model type&Model parameters&$\chi^2/n$&L$_{\rm
model}$, $10^{41}$ erg\\
&($10^{-3}$ counts s$^{-1}$)&&&&s$^{-1}$ (0.1--2.4 keV)\\
\hline
NGC 379&$6.1 \pm 1.0$&{\sc mekal}, free abundance&$kT = 0.96 \pm
0.15$ keV, abundance = $0.04 \pm 0.03$&8.6/7&3.0\\
&&power-law&$\alpha = 1.6 \pm 0.3$&12.1/8&6.6\\[2pt]
NGC 380&$24.2 \pm 1.2$&{\sc mekal}, free absorbing column&$kT = 0.59
\pm 0.02$ keV, $N_{\rm H} = (19 \pm 3) \times 10^{20}$ cm$^{-2}$&42.9/33&12.2\\[2pt]
NGC 383&$54.3 \pm 1.5$&{\sc mekal} + power-law (32\%)&$kT = 0.67 \pm
0.02$ keV, $\alpha = 0.53 \pm 0.08$&113/71&16.4\\[2pt]
NGC 382&$3.4 \pm 1.0$&{\sc mekal}, free abundance&$kT = 0.72 \pm 0.2$ keV,
abundance $< 0.02$&9.5/7&1.8\\
&&power-law&$\alpha = 1.9 \pm 0.3$&10.1/8&4.7\\[2pt]
NGC 385&$4.1 \pm 1.0$&{\sc mekal} + power-law (39\%)&$kT = 0.4 \pm
0.3$ keV,
$\alpha = 0.2 \pm 0.6$&4.1/5&2.1\\[2pt]
NGC 384&$2.9 \pm 1.0$&power-law&$\alpha = 1.1 \pm 0.4$&5.7/3&2.1\\
\hline
\end{tabular}
\vskip 5pt
\begin{minipage}{\linewidth}
Luminosities are unabsorbed, and are calculated assuming that all
objects are at the assumed distance of NGC 383/3C\,31 (luminosity
distance 73.0 Mpc); they are derived from 12 arcsec (4.1 kpc) source
circles. We use the {\it ROSAT} energy band to aid comparison with the
results of Komossa \& B\"ohringer (1999). Unless otherwise stated,
abundances used are 0.2 solar and the absorbing column is Galactic. For
multi-component models, the percentage quoted for the power-law
component is the fraction of counts between 0.5 and 7.0 keV
contributed by the model.
\end{minipage}
\end{table*}

Although the luminosities derived for the companion objects are
somewhat model-dependent, they are in broad agreement with the {\it
ROSAT}-derived values of Trussoni \etal\ (1997) and Komossa \&
B\"ohringer (1999). The luminosities are all in the general range
expected for isolated coronae of early-type galaxies (Forman, Jones \&
Tucker 1985) but several of the spectra are either best fitted with
non-thermal models or have a non-thermal component. This may indicate
a contribution from discrete X-ray sources, such as binaries, in these
galaxies, though the required luminosities ($\sim 10^{41}$ ergs
s$^{-1}$) are rather higher than expected for normal binary numbers.
In the case of NGC 383 the non-thermal component is attributed to the
AGN (Section \ref{core}).

\section{1E 0104+3153}

\begin{figure*}
\epsfxsize 10cm
\epsfbox{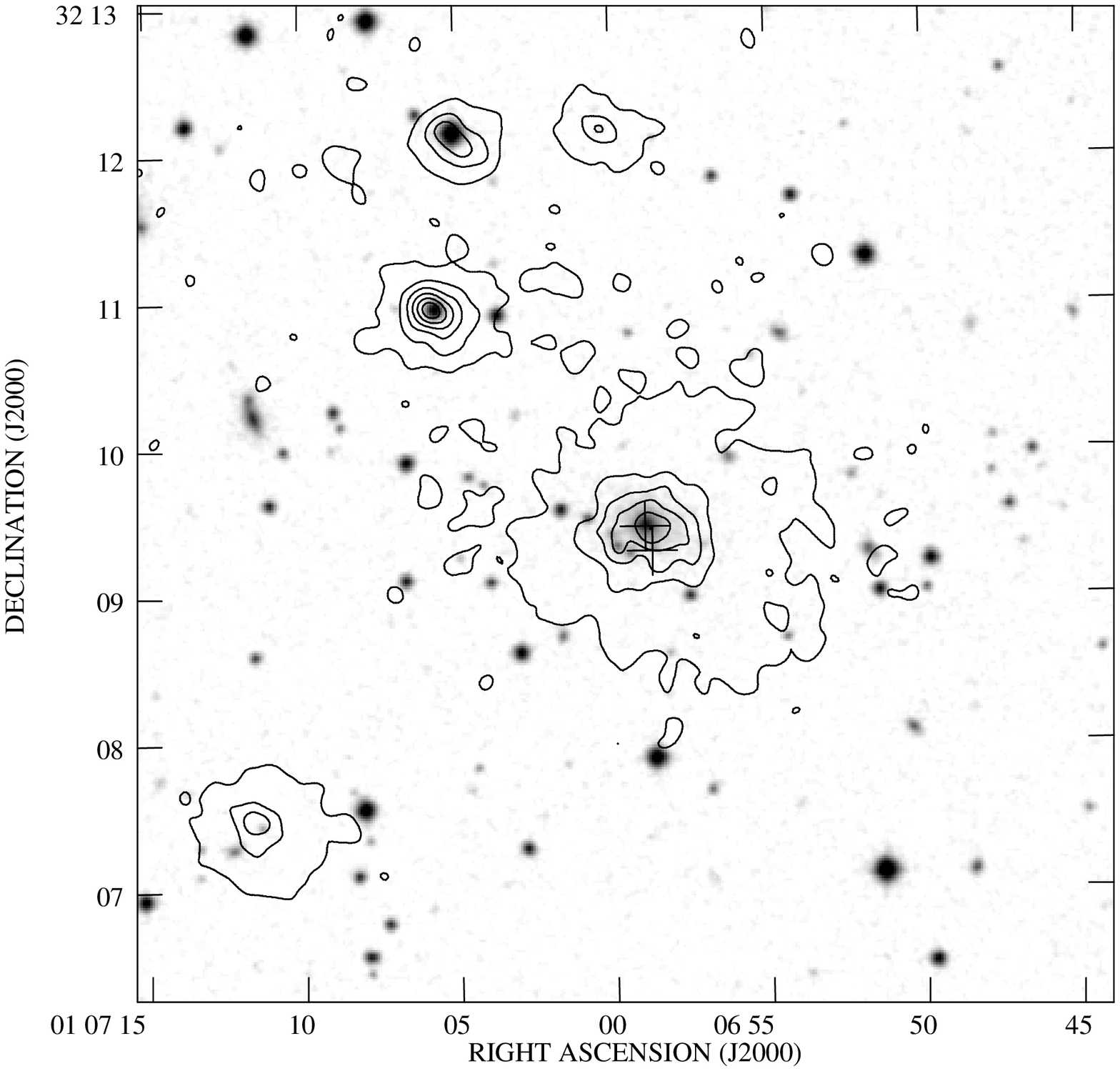}
\caption{DSS image of the region around 1E 0104+3153. Overlaid are
contours of the {\it Chandra} image, smoothed with a Gaussian kernel
with a FWHM of 8 arcsec; contour levels are at $0.04 \times (4, 10,
15, 20\dots)$ counts arcsec$^{-2}$. The lowest contour is at the
$3\sigma$ level, determined as specified by Hardcastle (2000); 1E
0104+3153 appears significantly more extended than other X-ray
detected objects in the field. The two X-ray sources to the NE appear
to be identified with stars. Crosses mark the positions of the
elliptical galaxy (northern object) and the BAL quasar, given by
Stocke \etal\ (1984).}
\label{1emap}
\end{figure*}

\begin{figure}
\epsfxsize \linewidth
\epsfbox{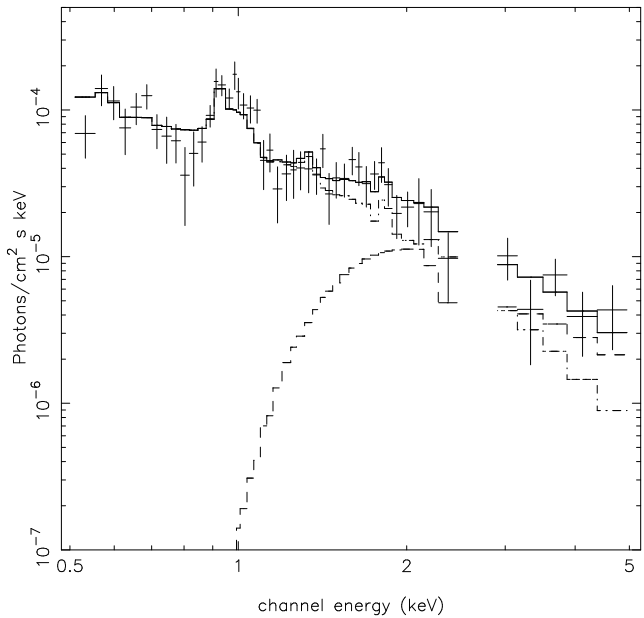}
\caption{The spectrum of 1E 0104+3153 with the best-fitting model,
consisting of a {\sc mekal} model and heavily absorbed power-law
component, as described in the text. The data at around 2.7 keV and
above 5 keV were excluded from the fit because these bins contained
few counts after background subtraction.}
\label{xspec-1e}
\end{figure}

The {\it Einstein} source 1E 0104+3153 lies 16.6 arcmin to the SW of
3C\,31, and is detected on the back-illuminated ACIS-S1 chip. The
X-ray source is close on the sky both to a $z=2.2$ broad
absorption-line (BAL) quasar and to a $z=0.11$ elliptical galaxy in a
group (Stocke \etal\ 1984; Fig.\ \ref{1emap}) and so the origin of the
X-ray emission is not clear. Komossa \& B\"ohringer (1999) obtained a
more accurate position for the X-ray source with the {\it ROSAT} HRI,
but could not resolve the ambiguity in the identification. At this
distance off-axis, the {\it Chandra} PSF is broad, with a half-energy
radius of $\sim 15$ arcsec. The spatial resolution of {\it Chandra}
therefore does not allow us to identify the X-ray source; in fact, the
{\it Chandra} position of the source centroid, at (J2000) RA
$01\hh06\mm58\fs737$, Dec.\ $+32\degr09'25\farcs15$, lies in between
the optical positions of the BAL object and the galaxy, which are RA
$01\hh06\mm58\fs756$, Dec.\ $+32\degr09'18\farcs03$ and RA
$01\hh06\mm58\fs978$, Dec.\ $+32\degr09'28\farcs02$ respectively
(Stocke \etal\ 1984), at 7 and 4 arcsec away from the centroid. The
source looks more extended than two nearby bright sources on the same
chip, which would tend to favour an identification with the emission
from the $z=0.11$ group. The large number of counts in the {\it
Chandra} observation, $1834 \pm 82$ in a 1-arcmin source circle with
background between 1 arcmin and 75 arcsec, allow us to make a good
determination of the source spectrum. A power-law model with
absorption fixed to the Galactic value gives a poor fit ($\chi^2/n =
93/47$, $\alpha = 0.7$). With excess absorption the fit is
significantly improved ($\chi^2/n = 80/46$, $N_H = 2 \times 10^{21}$
cm$^{-2}$, $\alpha = 1.3 \pm 0.2$), but there are still clear
residuals at around 1 keV. The data are also consistent with a {\sc
mekal} model at the redshift of the elliptical galaxy, with Galactic
absorption and fixed abundance 0.35; in this case we obtain $kT = 2.5
\pm 0.3$ keV, with $\chi^2/n = 56/47$. However, the best fits to the
data are obtained with a combination of a {\sc mekal} model at
$z=0.11$ and an absorbed power-law; we obtain $\chi^2/n = 39/44$ with
Galactic absorption, a temperature $kT = 1.8 \pm 0.1$ keV for the
thermal component, and for the absorbed power law a column of $(6 \pm
4) \times 10^{23}$ cm$^{-2}$ at the redshift of the quasar and $\alpha
= 1.9 \pm 1.3$. In this model (Fig.\ \ref{xspec-1e}), the thermal
plasma from the group contributes almost all the soft X-ray emission,
while the BAL power-law contributes to the high-energy tail, giving
rise to $\sim 10$ per cent of the total counts between 0.5 and 5 keV.
Such a high column density is not unexpected for a BAL object (e.g.,
Green \etal\ 1995) and the unabsorbed rest-frame 2--10 keV flux would
be $4 \times 10^{-13}$ ergs cm$^{-2}$, leading to a luminosity of $2
\times 10^{46}$ ergs s$^{-1}$, not implausible for a quasar. The
temperature of the thermal component is consistent with that measured
by Komossa \& B\"ohringer, and its luminosity in our extraction
region, assuming $z=0.11$, is $\sim 10^{43}$ ergs s$^{-1}$, which is
consistent with the temperature-luminosity relation for groups (e.g.,
Worrall \& Birkinshaw 2000). The statistics of the X-ray observation,
given the large {\it Chandra} PSF at this off-axis radius, are not
good enough to distinguish a significant offset between the centroids
of the X-ray source in the soft and hard bands, so we cannot be
certain that the quasar contributes to the hard emission. But, given
the extended appearance of the source and the lack of X-ray
variability reported by Komossa \& B\"ohringer (1999), we suggest that
the {\it Chandra}-band X-ray emission from this source is dominated by
the hot gas in the $z=0.11$ group.

\section{The extended emission}
\label{gas}

The arcmin-scale extended emission discussed by Trussoni \etal\ (1997)
and Komossa \& B\"ohringer (1999) is visible in our data mainly as an
increased background around 3C\,31. However, there is also a clear
extended halo around the core and jet of 3C\,31 (Fig.\ \ref{hijet}),
which is detectable by eye out to about 30 arcsec from the nucleus,
though the bulk of the counts lie within 10 arcsec. The 0.4--5.0 keV
luminosity of this component within 30 arcsec (10 kpc) is around
10$^{41}$ ergs s$^{-1}$, so that it is comparable to the galaxy-scale
gas components seen with {\it Chandra} around other FRI radio galaxies
(Worrall \etal\ 2001).

\begin{figure}
\epsfxsize \linewidth
\epsfbox{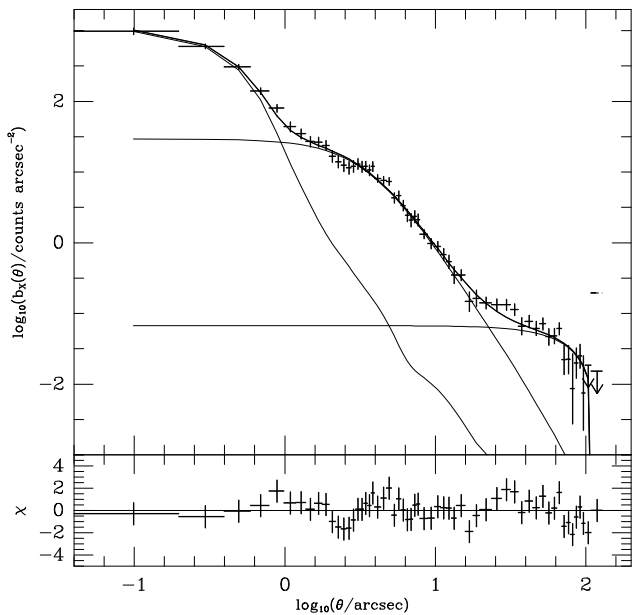}
\caption{The radial profile of the extended emission around 3C\,31,
with the best-fitting model as described in the text.}
\label{profile}
\end{figure}

To characterize the extended emission spatially we used software based
on the {\sc funtools} package to extract a radial profile directly
from the 0.5--7.0 keV events dataset. The profile was centred on the
nucleus, and excluded all detected background point sources, as well
as the pie slice between position angles of 320 and 360 degrees (to
avoid contamination from jet emission). The radial profile extended
out to 110 arcsec from the core, with background being taken between
110 and 130 arcsec; the size of the radial-profile region was
restricted by the requirement that it lie only on the S3 chip. Small
changes in the {\it Chandra} PSF on these angular scales do not
significantly affect our analysis.  Exposure correction was applied to
the radial-profile data using an exposure map calculated at the peak
energy of the dataset, 0.83 keV, as described in the {\sc ciao} 2.1
`science threads'. The resulting radial profile (Fig.\ \ref{profile})
was fitted with a model consisting of a point source (we use the PSF
parametrization of Worrall \etal\ 2001), a free $\beta$ model, and a
$\beta$ model with $\beta = 0.38$ and core radius 154 arcsec,
corresponding to the fit of Komossa \& B\"ohringer (1999) to the PSPC
data. [We choose to follow Komossa \& B\"ohringer rather than Trussoni
\etal\ (1997) because the former authors used an improved exposure
correction in the PSPC analysis (Feretti, private communication).]
Small changes in the {\it Chandra} PSF over the off-axis distances
involved do not significantly affect our analysis. The best-fitting
parameters of the free $\beta$ model were $\beta = 0.73 \pm 0.07$,
$r_c = 3.6 \pm 0.7$ arcsec (errors are $1\sigma$ for 2 interesting
parameters), with $\chi^2/n = 58.37/55$.  The unresolved core
component contained $761\pm 10$ counts and the central normalization
of the inner $\beta$ model was $33 \pm 3$ counts arcsec$^{-2}$, while
the larger $\beta$ model had a central normalization of $0.25 \pm
0.02$ counts arcsec$^{-2}$.

The central count density of the $\beta$ model means that we would
expect to see no more than around 200 counts from extended thermal
emission in the 1.5-arcsec extraction radius around the core discussed
in section \ref{core}, whereas the spectral fits to the core implied
that there was nearly twice this amount of thermal emission. These
facts can be reconciled if we hypothesize that there is an unresolved,
dense component of thermal emission close to the active nucleus. This
additional component, if present, does not affect the
dynamical analysis of Laing \& Bridle (2002b), which starts at a distance
of 2.5\,arcsec from the nucleus, and so we neglect it in what follows.

\begin{table}
\caption{Best-fitting temperatures and abundances for regions of
3C\,31's small-scale thermal emission}
\label{extfits}
\begin{tabular}{lrrrr}
\hline
Region&$kT$&Abundance&$\chi^2/n$&Deprojected\\
&(keV)&&&$kT$ (keV)\\
\hline
2.5--15 arcsec&$0.75 \pm 0.04$&$0.16 \pm 0.04$&46.6/33&--\\[4pt]
2.5--5 arcsec&$0.63 \pm 0.03$&$0.12 \pm 0.06$&18.5/13&$0.60 \pm 0.04$\\
5--10 arcsec&$0.78 \pm 0.05$&$0.15 \pm 0.06$&6.7/9&$0.76 \pm 0.05$\\
10--20 arcsec&$1.3 \pm 0.3$&$0.13 \pm 0.1$&0.1/3&$1.1 \pm 0.2$\\
\hline
\end{tabular}
\end{table}

We initially fitted a single thermal model to the spectrum of the
extended emission between 2.5 and 15 arcsec, using as background an
annulus between 20 and 30 arcsec from the core and excluding the
quadrant containing the jet. This region contains $820 \pm 37$ net
counts. Its spectrum is adequately fitted using a {\sc mekal} model with
Galactic absorption and free abundance, as shown in Table
\ref{extfits}. The temperature of $0.75 \pm 0.04$ keV is in good
agreement, within the errors, with the temperature of $\sim 0.6$ keV
estimated for the small-scale thermal component by Komossa \&
B\"ohringer (1999).

However, when we divided this region into three smaller annuli, we
found significant differences between the temperatures of the three
regions (Table \ref{extfits}). There appears to be good evidence for a
temperature gradient in this small-scale thermal emission, although the
best-fitting abundances are consistent. Assuming that there is no
abundance gradient, we then used the $\beta$ model fit to carry out a
simple deprojection of the three annuli, fitting three temperatures to
the inner annulus, two to the middle annulus and one to the outer
annulus, and scaling the emission measures of the extra components in
the inner annuli suitably to take account of the spatial variation as
parametrized by the $\beta$ model. We obtained good fits with an
abundance of $0.19 \pm 0.04$ and deprojected temperatures as tabulated
in Table \ref{extfits}. We adopt these temperatures, together with the
temperature $kT = 1.5 \pm 0.1$ keV obtained from the {\it ROSAT} data
at larger radii by Komossa \& B\"ohringer (1999), to parametrize the
temperature gradient in the thermal emission, fitting a model of the
form

\[
kT(r) = 
\cases{kT_0 + c r&$r < r_m$\cr
kT_L&$r>r_m$}
\]
where we take $kT_L = 1.5$ keV. For $r$ in arcsec, the best-fitting
values are $c = 0.047 \pm 0.015$ keV arcsec$^{-1}$ and $kT_0 =0.42 \pm
0.11$ keV (since the slope and intercept are strongly correlated, the
errors are $1\sigma$ for two interesting parameters). For the
best-fitting values we infer $r_m = 22.8$ arcsec.

\begin{figure*}
\epsfxsize 17.7cm
\epsfbox{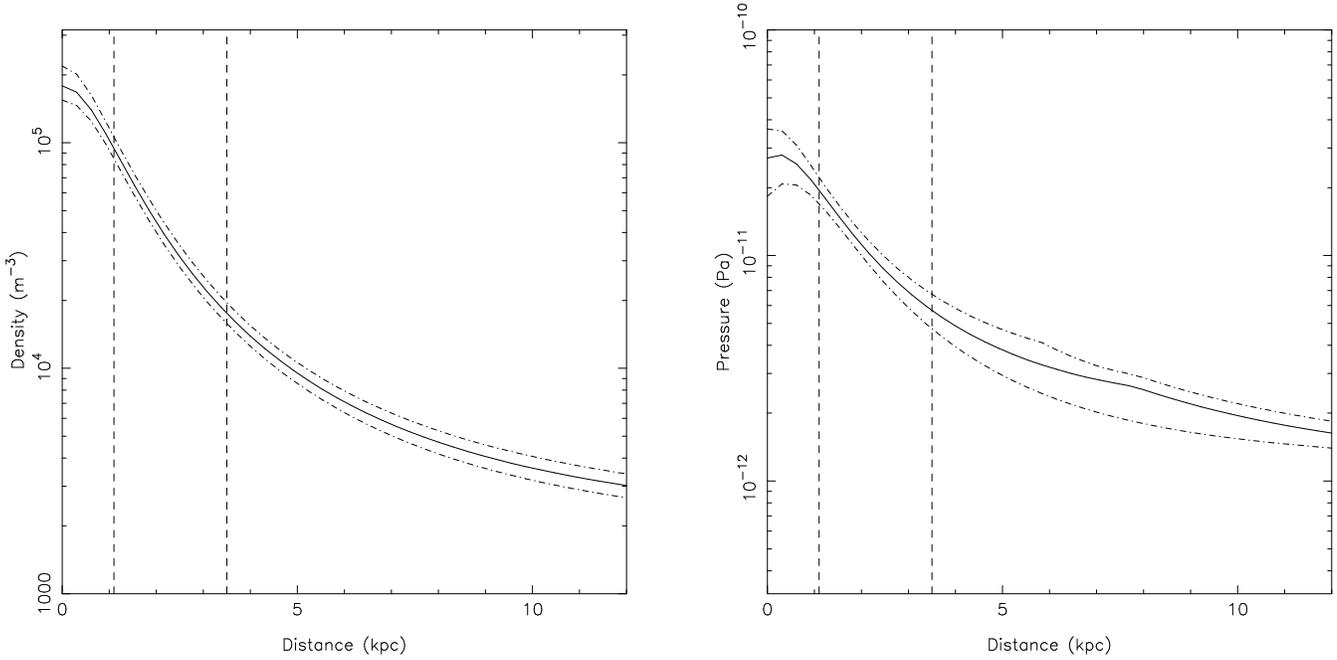}
\caption{Proton density (left) and pressure (right) as a function of
radius in 3C\,31, based on the model described in the text. The solid
black lines show the density and pressure derived from the
best-fitting $\beta$ models and the best-fitting straight line fit to
the temperature data of Table \ref{extfits}. The surrounding dotted
lines show the combined $1\sigma$ uncertainties due to the conversion
between central normalization and density, the uncertainties on the
$\beta$ model fits and (in the case of pressure) the uncertainties on
the linear fit to the temperature gradient. The slight discontinuity
in the pressure gradient at $r \approx 8$ kpc arises because of the
discontinuity in the simple temperature model we use. Because of the
possible additional thermal component discussed in Section \ref{gas},
the density and pressure may be underestimated in the inner 0.5 kpc.
The vertical dotted lines show the boundaries of the jet regions
defined by Laing \& Bridle (2002a); the inner region (0 to 1.1 kpc),
the flaring region (1.1 to 3.5 kpc) and the outer region (3.5 to 12 kpc).}
\label{pressure}
\end{figure*}

The relationship between volume-weighted emission measure and {\it
Chandra} counts in the 0.5--7.0 keV energy band is not a strong
function of temperature at temperatures around 1 keV, though it is
quite sensitive to abundance. Since we see no evidence for an
abundance gradient, we can safely assume that there is no strong
radial dependence of the count density on spectral parameters; it
therefore simply traces the density of X-ray-emitting gas, and the
analysis of Birkinshaw \& Worrall (1993) applies. Making the
approximation that the count density as a function of radius can be
written as the {\it sum} of the count densities from the two $\beta$
models involved, we use the results of Birkinshaw \& Worrall to
determine the proton density and, by using the observed temperatures,
the pressure as a function of radius in the small-scale X-ray-emitting
gas. In Fig.\ \ref{pressure} we plot the inferred density and
pressure. We tabulate the properties of the two $\beta$ models in
Table \ref{betas}.

\begin{figure}
\epsfxsize 8.3cm
\epsfbox{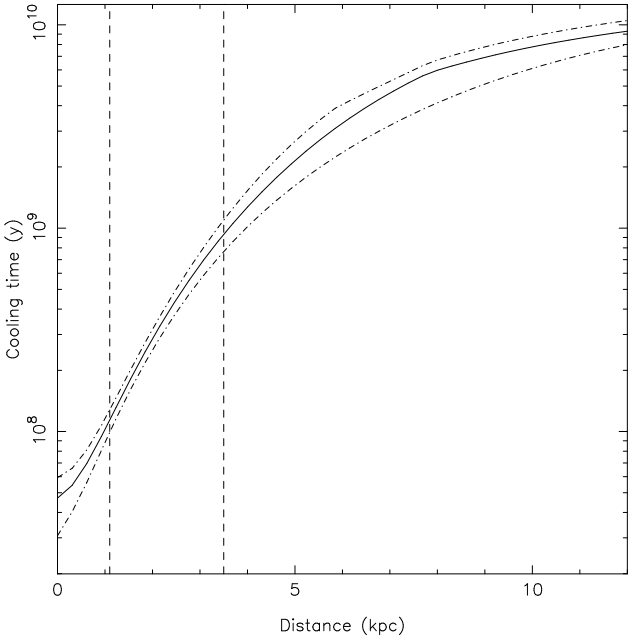}
\caption{Cooling time as a function of radius in 3C\,31. Lines and models as in
Fig. \ref{pressure}.}
\label{cooling}
\end{figure}

Fig.\ \ref{cooling} shows the inferred cooling time, calculated using
the {\sc mekal} model to take account of the fact that line cooling is
important at these temperatures. At $r=0$ the cooling time is only $5
\times 10^7$ years, and it is less than the Hubble time for $r \la 35$
arcsec (12 kpc). The unresolved thermal component discussed above
would be expected to have an even shorter cooling time. The central
regions of the thermal environment of 3C\,31 are therefore expected to
be cooling significantly, an expectation which is qualitatively
consistent with the observed decrease in temperature with decreasing
radius. Short cooling times are often inferred on kpc scales in
elliptical galaxies, but it is interesting that the cooling times we
derive for 3C\,31 mean that the central, compact, high-pressure
component, which Laing \& Bridle (2002b) argue is required to give the
radio jet its observed structure, is transient on time-scales
comparable to the lifetime of a radio source (and also to its own
sound-crossing time). Generalizing the model of Laing \& Bridle to
other twin-jet sources, we might expect that their host galaxies would
also contain dense central gas with a short cooling time. The {\it
causal} nature of the relationship between the presence of this gas
component and the the existence of a well-collimated jet is not clear
-- for example, we do not know whether there is a class of elliptical
galaxies without a central hot-gas component in which jets form but
are not recollimated and so do not give rise to strong radio emission.
It is also not clear whether (or by what mechanism) the radio source
can inject energy into the cooling gas and so maintain the high
pressure gradient required for recollimation, though we note that the
bolometric X-ray luminosity of the component represented by the
smaller $\beta$ model is a small fraction ($\sim 10^{-3}$) of the jet
energy flux estimated by Laing \& Bridle (2002b), so that the jet has
energy to spare for such a mechanism. High-quality observations of
more FRI radio sources, as well as of radio-quiet ellipticals, are
required to answer these questions.

\begin{table}
\caption{Properties of the $\beta$ models fit to the data}
\label{betas}
\begin{tabular}{lrrr}
\hline
Model&$\beta$&$r_c$ (arcsec)&Central normalization (m$^{-3}$)\\
\hline
Inner&0.73&3.6&$1.8 \times 10^5$\\
Outer&0.38&154&$1.9 \times 10^3$\\
\hline
\end{tabular}
\end{table}

\section{Conclusion}

We have used {\it Chandra} to observe the X-ray emission from the
nucleus and the jet of 3C\,31, and from hot gas in the inner regions
of its group environment, as well as making X-ray detections of the
other galaxies in the Arp 331 chain and of the nearby source 1E
0104+3153.

The non-thermal component of the X-ray core of 3C\,31 is well modelled
as a weakly absorbed power law, with no evidence for a more heavily
absorbed hard component. We have argued that this implies that the
X-ray emission originates outside any nuclear absorbing material.
The power-law spectrum of 3C\,31's core is relatively flat, and may be
consistent with an inverse-Compton origin for the X-ray emission.

The X-ray flux density and spectrum of the jet of 3C\,31 are well
constrained. Because of its steep spectrum and because (in spite of
uncertainties in the optical flux density) we can construct models
which connect the radio and optical data smoothly to the X-ray, we
attribute the X-ray emission from the jet to the synchrotron process.
We know of no FRI radio source with a bright radio jet which does {\it
not} have an X-ray jet, plausibly of synchrotron origin, when imaged
deeply in the X-ray. It may be that the required high-energy, {\it in
situ} particle acceleration is universal in these objects. The inner
part of 3C\,31's jet has a higher ratio of X-ray to radio emission
than the rest of the jet. This is seen in a number of other sources
(e.g., 3C\,66B, Hardcastle \etal\ 2001; Cen A, Kraft \etal\ 2002) and
may indicate a difference in the particle acceleration mechanism in
this region.

Our observations have resolved for the first time the regions of
3C\,31's hot-gas atmosphere where the radio jets are known to flare and
decelerate (Laing \& Bridle 2002a). There is significant evidence for
a temperature gradient as well as a strong emissivity gradient as a
function of radius. We have determined the density and pressure in
this region as a function of radius, and have shown
that the cooling time of this central hot-gas component is short. A
companion paper (Laing \& Bridle 2002b) discusses the use of the
density and temperature profiles to constrain the physical parameters
of the jets.

Our luminosity measurements for the other galaxies of the Arp 331
chain are consistent with those of other workers, and in most cases
the observed luminosities and spectra are best explained as
originating in small-scale thermal emission associated with the
galaxy. In the background source 1E 0104+3153, whose origin has long
been uncertain, we infer, from the spatial and spectral properties of
the X-ray source, that almost all the soft X-ray emission originates
in the $z=0.11$ group rather than the known BAL quasar close to the
X-ray centroid. The quasar may contribute to the observed hard X-rays
given a plausible intrinsic absorbing column.

\section*{Acknowledgments}

We are grateful to Judith Croston for providing optical flux densities
for the 3C\,31 jet region in advance of publication.

The National Radio Astronomy Observatory is a facility of the National 
Science Foundation operated under cooperative agreement by Associated
Universities, Inc.

\clearpage

\end{document}